\documentclass{camera}

\newcommand{\be}{\begin{equation}}  
\newcommand{\ee}{\end{equation}}
\newcommand{\bea}{\begin{eqnarray}}  
\newcommand{\eea}{\end{eqnarray}}

\begin{document}

%
\title{Theory of neutrino masses and mixing}

%
\author{Alexei Y. Smirnov}

%
\organization{Max-Planck-Institut f\"{u}r Kernphysik, 
69029 Heidelberg, Germany, \\ 
and\\
International Centre for Theoretical Physics, 
34014 Trieste, Italy}

\maketitle

\begin{abstract}
In spite of enormous experimental progress 
in determination of the neutrino parameters, theory of neutrino mass and mixing 
is still on the cross-roads.  Guidelines could be (i) the connection between  
zero neutrino charges (and therefore a possibility to  be Majorana particle),  
smallness of the neutrino mass  and  large lepton mixing,  
(ii) joint description of leptons and quarks, (iii) 
existence of the right handed (RH) neutrinos without special quantum numbers. 
Properties of the RH neutrinos and the UV completion of the seesaw 
may turn out to be the key to understand the neutrino mass and mixing.  
In view of the LHC results minimalistic scenarios like $\nu$MSM  
look rather plausible. 
Still the GUT's with additional hidden sector, QLC, 
high scale flavor symmetries  are appealing.  
Concerning mixing, the main issue is  ``symmetry or no symmetry''  
behind the observed pattern. The symmetry group condition is useful tool 
to study consequences of symmetries and  
to perform ``symmetry building''. Sterile neutrinos are challenge 
but  also opportunity for the present theoretical constructions.

\end{abstract}

%

\section{Introduction}

The title of this talk~\footnote{Talk given at the Pontecorvo 100 Symposium
Pisa, Italy, September 18 - 20, 2013.} is a joke 
in a spirit of Bruno Pontecorvo. 
In reality  
we have plenty of  observations, 
mechanisms, schemes, models, 
approaches, conjectures and even scans of various possibilities: 
symmetries, parameters, field contents (see reviews ~\cite{reviews}). 

It is widely accepted (which does not mean much) that new physics  
beyond the standard model and beyond just  adding 
of the right handed neutrinos is involved. 
However, the  proposed mass scales of this new physics range 
from eV to the Planck mass, that is, 28 orders of magnitude.  
For explanation of mixing  a spectrum of ideas spans from symmetry  
to anarchy~\cite{anarchy}.  This means that  we are far from real 
understanding of the underlying physics.  
As a consequence, we can not predict unambiguously mass hierarchy, 
the value of  CP  phase, the absolute scale of neutrino mass, {\it etc.}.  

Frameworks under discussion cover   
minimalistic phenomenological scenario of $\nu$MSM 
on one extreme,   and sophisticated structures 
at different energy scales on another. 
Studies spread from simple-minded manipulations with mass and 
mixing matrices to consideration of geometric,  
strings  as well as complicated dynamics origins 
of the observed  pattern.   
Certain models and approaches can indeed correspond to  
reality, and still some key elements may  be missed.  

Recent trends are determined by the fact that 
no new physics has been discovered at LHC and other experiments.  
Higgs boson properties are in agreement with those of the SM. 
This hints that we should be  stingy in our speculations, 
and take more seriously scenarios with nothing or almost nothing
up to the Planck scale. 

In this connection 
some  ideas, approaches and  results will be reviewed  
which may have a chance to reflect reality.  
I will consider problems of construction 
of the theory of neutrino mass and mixing.

\section{Facts and feelings} 

The data from now numerous solar, atmospheric, reactor, accelerator 
neutrino experiments can be nicely described in the $3\nu$ mixing  
framework. All the mixing angles, $\theta_{ij}$,   
as well as  mass squared differences $\Delta m^2_{21}$ and  
$|\Delta m^2_{31}|$  are determined with 
rather good accuracy \cite{global}.  
As far as theory is concerned, even after precise measurements 
of the 1-3 mixing  we are at the cross-roads. 
The same value of 1-3 mixing has  different relations 
to other parameters with completely different implications. 
The most appealing possibilities are  

1. ``Naturalness'' - absence of fine tuning in  the mass matrix 
gives \cite{zhenia}: 
\be
\sin^2 \theta_{13} =   
A \frac{\Delta m_{21}^2}{\Delta m_{31}^2}, ~~~~~  
A = {\cal O}(1).  
\ee

2. Connection to Cabibbo mixing \cite{tanimoto}: 
\be
\sin^2 \theta_{13} \approx  \frac{1}{2}\sin^2 \theta_{C},  
\ee
which can be realized in the context of Quark-Lepton Complementarity 
\cite{qlc} with 
implications of GUT or/and family symmetry \cite{raidal}.    

3. Connection to deviation of 2-3 mixing from maximal:  
\be
\sin^2 \theta_{13} \approx  \frac{1}{2}\cos^2 2\theta_{23} ~~
{\rm or}~~  \theta_{13} \approx  \sqrt{2} (\pi/4 - \theta_{23}),  
\label{sin13}
\ee
which was predicted in model with   $T^{\prime}$ symmetry
\cite{frampton} but may also follow from the universal 
$\nu_\mu - \nu_\tau$ symmetry violation \cite{myrev}.    

4. Inter-generation connection: 
\be
\sin^2 \theta_{13} 
\approx \frac{1}{4} \sin^2 \theta_{12}\sin^2 \theta_{23},  
\ee
which is analogous to the quark relation 
$V_{ub} = 0.5 V_{us} V_{cb}$ (the $q-l$ similarity). 
This may follow from a kind of Fritzsch texture for 
mass matrices (with texture zeros, U(1) symmetry, etc.).
For more cases see \cite{myrev}.  



Determination of other unknowns  (mass ordering,  CP violation phase, 
absolute scale of mass) may or may not  help 
depending on the outcome. 
E.g., in the case of normal mass ordering possible    
hierarchy is weak or absent: 
$m_2/m_3 \geq \sqrt{\Delta m_{21}^2 / \Delta m_{31}^2 }$. 
Still this resembles the situation in quark sector. 
Inverted mass hierarchy  
implies strong degeneracy:   $\Delta m_{21}/m_1 \leq 1.6 \cdot 10^{-2}$,  
and therefore certain symmetry. 
Even more symmetry will be realized if the spectrum is degenerate.  
There are first glimpses of the CP phase: 
global fits (essentially the atmospheric neutrino data) 
indicate $\delta_{CP} \sim \pi$ \cite{global}.

In view of the present  trends 
it makes sense {\it to refine and sharpen known arguments,}  
as actually Pontecorvo did continuously. 
There are three well established facts about neutrinos: 
\be
(i)~~Q_\gamma = Q_c = 0, ~~~~~ (ii)~~ m_\nu \ll m_l, m_q, ~~~~~ 
(iii) ~~  \theta_{12} \sim \theta_{23} \sim 1
\label{facts}
\ee
and zero values of conserved charges (i)  mean  that neutrinos can be 
Majorana particles.  Naturally, one expects that 
these facts are  connected. 

{\it SM and Weinberg operator.}  
After decoupling of possible heavy degrees of freedom 
of new physics one obtains in  
the lowest order the D=5 operator~\cite{weinberg} 
\be
\frac{1}{\Lambda} L L H H,  
\label{d5}
\ee
where $\Lambda$ is the scale of new physics. 
The operator generates Majorana neutrino mass 
and to a large extend, realizes the connection (\ref{facts}). 
Is this the end of ``theory of neutrino mass and mixing''? 
Can we say more and advance further?  
As {\it guidelines and prejudices} I would take  

\begin{itemize}

\item 
Minimality and connection (\ref{facts}).  

\item
Quark-lepton analogy 
(correspondence, symmetry, unification). 
Theory of masses and mixing should include 
both quarks and leptons,  
although some new elements may be present 
in lepton sector. 

\item 
Existence of the right-handed neutrinos
{\it without} any special symmetry (new quantum numbers).

\end{itemize}

\section{RH neutrinos, Seesaw and its UV-completion.}   

Once the RH neutrino components without special quantum numbers 
are introduced,  unavoidably  neutrinos should  have the Dirac mass terms as all 
other leptons and quarks have.  Then smallness of the mass can be due to 
very small Yukawa couplings (still to be explained) or  
some new physics.
The natural way to suppress the neutrino Dirac 
mass is to introduce large  Majorana masses of   
$\nu_R$, realizing the seesaw~\cite{seesaw}: 
$m_\nu = - m_D M_R^{-1} m_D^T$. In minimal 
version the seesaw simultaneously suppresses 
the Dirac mass and generates small  Majorana mass 
without introduction of new symmetry. 
However, seesaw may provide mainly suppression 
of the Dirac masses,  if e.g. the RH neutrino masses are at the Planck scale.  
Then dominant contribution to $m_\nu$ comes from another mechanism. 

In the minimal version of seesaw the scale 
of $\nu_R$ masses is 
\be
M_R \sim \frac{m_D^2}{m_\nu} \sim 10^{14}  
~{\rm GeV}.  
\label{rhscale}
\ee 
That is,  $\nu_R$ introduces new mass scale which is 
much smaller than $M_{PL}$.  
Smallness of neutrino mass is another indication of existence 
of new physics scale  apart from  unification of gauge couplings.

The existence of heavy $\nu_R$ affects the Higgs sector 
in two ways: 

1. It gives (the loop with $\nu_l$ and $\nu_R$) 
the correction to the Higgs mass which is 
quadratically divergent~\cite{vissani}:   
\be
\delta m_H^2 \approx \frac{y^2}{(2 \pi)^2} M_R^2 
{\rm log} (q/M_R) 
\sim \frac{M_R^3 m_\nu}{(2 \pi v)^2} ~ {\rm log} (q/M_R),  
\label{deltamh}
\ee
where $v$ is the electroweak VEV and 
$y$ is the Yukawa coupling. Typically one gets 
$\delta m_H^2 \sim (10^{13}~ {\rm GeV} )^2$.

2. It modifies (loop with $\nu_l, \nu_R, \nu_l, \nu_R $) 
the  RG running of  the quartic Higgs coupling 
$\lambda$,  and consequently,  modifies the Higgs potential.  
The high scale minimum becomes deeper 
which in turn,  affects  stability (lifetime) 
of the EW vacuum \cite{stabil}. For Higgs masses below 128 GeV 
effect is small.  


Possible solutions of the  problem  1) are  

- Cancellation of contribution  (\ref{deltamh}) due to 
existence  of new particles  
(which is realized, e.g.,  in SUSY), 
or fine tuning, in which case dependence of low scale
observables, in particular Higgs mass, on high scale physics parameters 
(mass of RH neutrino) becomes enormous.  

- Reduction of the seesaw scale $M_R \sim M_{EW}$.  This requires small Yukawa
couplings or cancellation of contributions from different RH neutrinos.  

- Increase of the seesaw scale up to the Planck scale 
$M_R \sim M_{Pl}$   
by the prize of introduction of e.g. many RH neutrinos. 
In this case one can simply ``blame'' 
some new Planck scale physics which 
is responsible for tuning of parameters. 

- Modification of the seesaw mechanism: 
introduction of  more degrees of freedom,  
in particular, more than 3 RH neutrinos. 
In  the  double seesaw \cite{dseesaw}  three additional singlets $S$ 
with Majorana masses $\mu$ couple to (mix with) 
the RH neutrinos via the Dirac mass 
term $M_D$. Then  
\be 
m_\nu = m_D M_{D}^{-1} \mu M_{D}^{T-1} m_D^{T}. 
\ee
There are three possibilities depending on the size of the 
lepton violating mass $\mu$.  

(i) $\mu = 0$ which gives one massless neutrino  per generation. 
This is an example of multi-singlet (or ``chiral mismatch'') mechanism of  
suppression of the  Dirac mass.  
Physical consequences include  
violation of universality and unitarity for lights states  
characterized by the ratio $m_D/M_D$.  

(ii) $\mu \ll M_D$ corresponds to the  inverse seesaw.  
It allows to lower the scales of neutrino mass generation
and still have large enough probability 
of production of new heavy states at LHC.  
Spectrum of the heavy states is composed of pseudo-Dirac heavy 
leptons with small mass splitting. 
For $M_D \sim$ TeV, one has $\mu \sim $ kev 
which can be  generated radiatively \cite{dev}.  

(iii) $\mu \gg M_D $ -   
cascade seesaw.  This leads to  masses of the RH neutrinos 
$M_R \sim M_{D}^2/\mu$. 

The ultraviolet completion of the high scale 
seesaw  can be one of the driving forces of 
new developments. 
Interestingly, other mechanisms are also related to properties 
of the RH neutrinos. 
Thus, in the case of extra spatial dimensions \cite{add}, \cite{rs} 
the overlap mechanism works with  different 
localizations of the left and right handed neutrino components.  

\section{Large mass scales,  small mass scales  and LHC}

Low scale mechanisms of neutrino mass generation  can, in principle,  be tested 
at LHC and other laboratory  experiments.  Some   
possibilities include. 


1. Radiative mechanisms. 
The  main features of e.g. the 
one loop mechanism from  \cite{1loop} 
are (i)  absence of  usual RH neutrinos;
(ii) new Higgs doublet $(\eta^+ , \eta^0)$,  
(iii) fermionic singlets $N_k$. 
They are odd under discrete symmetry $Z_2$, 
whereas the SM particles are $Z_2$ even;    
$\eta^0$ has zero VEV. If $Z_2$ is exact, 
$\eta^0$ or the lightest  $N_k$ are stable 
and can be the Dark Matter particles. 
So, here a popular neutrino mass - DM connection is realized.

Two loops Zee-Babu mechanism and its  modifications
are still ``on the market'' \cite{babu}.  Models have no RH neutrinos, 
new scalar singlets $h^-$  and $k^{++}$ are introduced.  
The models are testable:  
new charged bosons can be produced at LHC, 
decays $\mu \rightarrow \gamma  e$, and  
$\tau \rightarrow 3 \mu$  are predicted with rates 
within  reach of  forthcoming experiments. 

2. Smallness of neutrino mass can be due to new Higgses,  
e.g.,  Higgs triplet in the seesaw 
type II  or new Higgs doublets  with small VEV's. 

3. Low scale L-R symmetry models with 
$M(W_R) \sim$ few TeV and $M(N_R) \sim$ 0.5 - few TeV 
\cite{LRmod}  have several different 
contributions to the light  neutrino masses, 
including the see-saw type I with  small Yukawa couplings, 
Higgs triplet mechanism.  Signatures of the models at LHC are 
the same-sign bi-leptons, 
$l l j j$, and no missing energy~\cite{goran}. 
For $M (W_R)  > M (N_R)$   
resonance production  of $W_R$ occurs.  
Peaks at the invariant mass  $W(jj l) = m_N$, and 
$W(jj ll) = m_W$ should be observed. 
In the t-channel the corresponding diagram 
coincides with  the diagram for $ 0\nu \beta\beta $ decay. 
Consequently, complementary bounds from  
LHC  and  $0\nu \beta\beta$ can be obtained \cite{lhcbb}.

The $\nu MSM$ scenario \cite{numsm} deserves now special attention. 
Its signature is that 
nothing should be seen at LHC. Everything is at or  below the EW scale
including masses of the RH neutrinos. Consequently,  
small Yukawa couplings should be introduced. 
Spectrum of the model consists of 
a) two strongly degenerate states with masses 
$\sim  (0.1 - 5)$ GeV, and splitting $(10^{-3} - 1)$ eV. 
They  generate light masses of neutrinos 
via seesaw, and the lepton 
asymmetry in the Universe via $\nu-$oscillations.  
They can   be produced in the $B-$meson decays 
with BR $\sim 10^{-10}$.  b) One RH neutrino with  mass (3 - 10) kev and very small 
mixing with active neutrinos plays the role of  warm dark 
matter. It can be searched for through its radiative decays 
as an X-ray line \cite{numsm}. 

The Higgs inflation scenario can be realized \cite{bezrukov} which 
is in a good agreement with the Planck data. 
Nothing new below Planck scale is expected.  
Among open questions are  smallness of the Dirac Yukawa couplings,  
huge coupling of Higgs to gravity, extremely small mass 
splitting between heavy RH states,  etc. 

High scale seesaw mechanisms can not be probed at LHC either.  
Here there are two interesting realizations: 1).  GUT seesaw 
with $M_R \sim M_{GUT} \sim 10^{16}$ GeV
which is possible for the heaviest 
RH neutrino.  
Leptogenesis is due to  the CP-violating  out of 
equilibrium decay of RH neutrinos.  
2). Double (cascade) seesaw with $\mu \sim M_{Pl}$ and  $M_D \sim M_{GUT}$  
explains the intermediate scale (\ref{rhscale}) 
for the RH neutrinos $M_R \sim M_{GUT}^2/M_{Pl} \sim 10^{14}$ GeV. 

An appealing scenario is the SO(10) GUT with 16-plet fermions, 
hidden sector at GUT - Planck scales composed of fermion and scalar  
singlets of SO(10).  
Presence of the fermion singlets with addition of the Higgs 16-plet 
can  realize high mass scale  double seesaw, 
enhance mixing, 
generate zero order mixing pattern, produce randomness (if needed).   
Flavor symmetries at high  (GUT, above GUT) scales 
can ensure specific form of the RH neutrino mass matrix,  
and consequently, specific form of $m_\nu$. 


\section{Mixing: symmetry or no symmetry}

The observed pattern of the lepton mixing  
can be described by the approximate TriBimaximal (TBM) mixing \cite{tbm}. 
Is a symmetry behind the mixing 
(usually non-abelian discrete symmetries are used 
\cite{discrete}) real or accidental?  
TBM looks strange in a sense that 
it is difficult, if possible, to  connect it to the lepton  masses  
although both masses and mixing result from diagonalization of 
the same mass matrices.  
Framework which could realize such a feature 
is that mixing originates from different ways of the flavor symmetry 
breaking in the neutrino and charged lepton (Yukawa) sectors \cite{framework}. 
These different ways lead to different 
residual symmetries of the mass matrices of neutrinos and charged leptons:
\be
G_f \rightarrow  
\left\{
\begin{array}{l}
G_\nu \\
G_l 
\end{array}. 
\right.
\ee
Furthermore, $G_\nu$ and $G_l$ should be generic symmetries 
which do not depend on values of masses. 
This ensures maximal control of mixing by symmetries.  

It has been shown that in this framework the mixing parameters 
or relations between them can 
be obtained without 
model-building, immediately from knowledge 
of the residual symmetries \cite{dani1, myrev}. 
Model-building is the ugliest  part of the construction: it requires    
many assumptions, {\it ad hoc} introduction  
new fields, auxiliary symmetries,  
new tuned parameters. 
One however can skip this ``unpleasant'' part 
and  obtain final results in one step even without 
construction of mass matrices:  
\be
residual~ symmetries~ \rightarrow 
~relations ~between ~mixing ~ parameters. 
\ee
This can be done using the symmetry group condition \cite{dani1}: 
\be
(U_{PMNS} S_i U_{PMNS}^{\dagger} T )^p  = I.     
\label{group}
\ee
Here $S_i$ and $T$ are the symmetry transformations of the 
diagonal neutrino  and charged lepton mass matrices
({\it i.e.} matrices in the mass bases)  and $p$ is an integer.  
In the flavor basis ($T$ is kept the same) 
the transformation  matrix of neutrinos becomes  
$S_{iU} = U_{PMNS} S_i U_{PMNS}^{\dagger}$. 
So, the relation (\ref{group}), which can be 
rewritten as $(S_{iU} T)^p = I$, 
is nothing but condition that  the elements 
$S_{iU}$ and  $T$ form a finite group. 
(The product  $W \equiv S_{iU} T$ also belongs to the group 
and due to finiteness the integer $p$ 
exists such that $W_i^p = I$)~\footnote 
{Notice that whole information about mixing is 
in the mass matrices only in the basis where 
the charged currents are diagonal. So,  the  
neutrino and charged lepton symmetry transformations 
should be taken in such a  basis.}.    

The relation (\ref{group}) connects the mixing 
matrix $U_{PMNS}$ and generating elements of the group 
in the mass bases. It is equivalent to 
\be
{\rm Tr} (U_{PMNS} S_i U_{PMNS}^{\dagger} T) = a, ~~~~    
a =  \sum_j \lambda_j, ~~~\lambda_j^p = 1,   
\label{grrel}
\ee
and $\lambda_j$ are  three eigenvalues of 
$W_{i}$~\cite{dani1}.  
The generic (mass independent) transformation matrices 
for neutrinos are 
$S_1 = diag (1, -1, -1)$ and $S_2 = diag (-1, 1, -1)$
which correspond to maximal generic symmetry 
${\bf Z}_2 \times {\bf Z}_2$,   
and for charged leptons one can take $T = 
diag \left(e^{i\phi_e}, e^{i\phi_\mu}, e^{i\phi_\tau}
\right)$, 
where  $\phi_\alpha = 2\pi \kappa_\alpha /m$ 
corresponds to ${\bf Z}_m$.  


For single fixed $S_i$  ($G_\nu = {\bf Z}_2$) the relation (\ref{grrel}) 
determines $i$th column of the mixing matrix. 
Maximal  $G_\nu = {\bf Z}_2 \times {\bf Z}_2$ fixes two columns 
and therefore determines the mixing matrix completely. 
The only phenomenologically viable structure here is TBM and so 
it seems  that indeed TBM is special. 
One can scan all the possibilities varying 
$p, \kappa_\alpha, m$ reproducing results of~\cite{Lim}. 
However,  only for special sets of these parameters 
the group is finite, see~\cite{lam-2}.  

The symmetry $G_\nu$ can  be extended by  transformations 
which left the neutrino mass matrix invariant only for 
specific mass spectra \cite{dani3}. 
In this case  relations will include also masses and  
Majorana CP phases. 
For unitary symmetry transformations  
the possibilities include 
a) Partially degenerate spectrum, 
$m_1  = m_2, ~m_3 $, and   
$S_\nu = O(2)$,  so that $G_\nu = SO(2) \times Z_2$.  
That leads to the relations \cite{dani3}
\be
\sin^2 2\theta_{23} = \pm \sin \delta = \cos \kappa = m_1/m_2 = 1
\label{massrel}
\ee
with  1-2 mixing being undefined. The relation 
(\ref{massrel}) can  
be a good first order approximation 
both for normal and inverted mass hierarchies. 
Relatively small corrections  to the mass matrix can lead to 
the required 1-2  mass splitting and mixing. 
b)  Degenerate  mass spectrum
$m_1 = m_2  = m_3$ \cite{dani3}. c). Spectrum with  one zero mass 
$m_1 = 0,  m_2,  m_3$, in which  case 
$G_\nu$ is a subgroup of $U(3)$ \cite{joshipura}. 

Other scenario is when flavor symmetry 
is broken down to the same residual symmetry 
in neutrino and lepton sectors or no symmetry is left.  
(The same flavon fields  are responsible for  
the neutrino and charged lepton mass generation.) 
In this case mixing  can originate from  
(i) different nature of the mass terms 
of the charged leptons (Dirac) and neutrinos (Majorana), 
(ii) mixing of neutrinos  
with new degrees of freedom $S$ - singlets of SM. 

\section{Remark on sterile neutrinos}

I am sure Bruno Maksimovich would enjoy knowing that 
a number of researchers working on his sterile neutrinos  got 
emails saying  ``Dear Dr.  ... ,
Please pay attention to our upcoming Special Issue on  
"Research in Sterility" which will be published in the
"Advances in Sexual Medicine",  an open access journal. 
We cordially invite you to submit your paper.''

``Steriles'' with mixing parameters, $U_{iS}$,  and mass $m_S$  
required by the  LSND/MiniBooNE,    reactor and gallium 
anomalies \cite{giunti} are non a small perturbation
of the $3\nu$ picture. In the presence of steriles 
the mass matrix of active neutrinos becomes~\cite{renata}  
\be
m_\nu =  m_a  + \delta m,   
\ee
where $m_a$ is the original active neutrino mass 
matrix which follows, e.g., from see-saw,
and $\delta m$ is the mass matrix 
induced by mixing with steriles. In the $(3 + 1)$ scheme 
$\delta m_{ij} \approx - 
U_{iS} U_{jS} m_{S}  \approx 0.04 m_{S}$. 
That is, $\delta m  \approx 0.04$ eV if $m_S \sim 1$ eV,  
which is comparable with the largest elements of $m_a$ 
for hierarchical mass spectrum. 
So, the correction $\delta m$ is not a perturbation,  
it can change structure (symmetries) of the original 
mass matrix completely: it can 
produce the dominant $\mu\tau$ - 
block with small determinant,  
enhance lepton mixing, generate TBM mixing,   
be, in general, the origin of difference of  $U_{PMNS}$ and $V_{CKM}$. 
Thus, checks  of existence of these steriles  
are of the highest priority  for 
further theoretical advances.

\section{Conclusions and outlook}

Theory of neutrino masses and mixing is still 
at the cross-roads with many possibilities.  
In particular,  the same value of 1-3 mixing satisfies various 
relations which have different implications.  
Interesting possibilities  include natural mass mixing relation,   
the QLC relation, special violation of the 
$\nu_\mu - \nu_\tau$ symmetry, quark-lepton similarity,   
{\it etc.}. The question ``Can we go beyond 
D5 Weinberg's operator'' is still open.

Elements of theory which have some chance to reflect reality probably include 
connection between zero charges, smallness of mass and large mixing, 
unified description of quarks and leptons, 
existence of the RH neutrinos without special quantum numbers.

Low scale mechanisms of $\nu$ mass generation are not much 
along the guidelines,  
but they are testable  and deprived  of the hierarchy problem. 
In view of data from LHC, MEG,  etc.,   
the minimalistic phenomenological scenario of $\nu$MSM  
looks more plausible than before. 
Still a  scenario with  high scale seesaw,  probably in some extended version 
(with more RH neutrinos involved),    
some flavor symmetry at the high mass scale,  
unification of quarks and leptons,  
similarity of the Dirac structures in both sectors looks appealing. 
The high scale seesaw creates the hierarchy problem and
influences stability of the Higgs potential. Solutions 
of these problems may lead to some new developments.
  
It look like RH neutrinos 
their existence or non-existence their number and properties 
are the key to understand mass and mixing of light neutrinos. 
Smallness of neutrino mass  may be 
connected to other hierarchies.

Concerning mixing pattern, the main issue is ``Symmetry or no symmetry''  
behind the observed pattern. 
TBM  could be accidental and symmetry behind - misleading 
in searches for underlying physics. 
As the zero order structure it is still possible, 
and still useful as book-keeping for  
phenomenological considerations.

The symmetry group relations are a powerful tool for 
studies of consequences of discrete flavor symmetries 
for lepton mixing and masses.  They are  useful for 
``symmetry building'':  uncovering symmetry for a given 
mixing pattern. 

Sterile neutrinos are challenge 
for the standard $3\nu$ scenario. Tests of existence of 
these sterile neutrinos are of paramount importance.


\begin{thebibliography}{99}

\bibitem{reviews}
G.~Altarelli and F.~Feruglio,
  New J.\ Phys.\  {\bf 6} (2004) 106; 
R.~N.~Mohapatra and A.~Y.~Smirnov,
  Ann.\ Rev.\ Nucl.\ Part.\ Sci.\  {\bf 56} (2006) 569;  
G.~Altarelli and F.~Feruglio,
  Rev.\ Mod.\ Phys.\  {\bf 82} (2010) 2701
 S.~F.~King and C.~Luhn,
  Rept.\ Prog.\ Phys.\  {\bf 76} (2013) 056201. 


\bibitem{anarchy}
  A.~de Gouvea and H.~Murayama,
  Phys.\ Lett.\ B {\bf 573} (2003) 94;  
  arXiv:1204.1249 [hep-ph].




\bibitem{global} 
  F.~Capozzi, G.~L.~Fogli, E.~Lisi, A.~Marrone, D.~Montanino and A.~Palazzo,
  arXiv:1312.2878 [hep-ph].


\bibitem{zhenia}
  E.~K.~Akhmedov, G.~C.~Branco and M.~N.~Rebelo,
  Phys.\ Rev.\ Lett.\  {\bf 84} (2000) 3535; 
  W.~Rodejohann, M.~Tanimoto and A.~Watanabe,
  Phys.\ Lett.\ B {\bf 710} (2012) 636. 

\bibitem{tanimoto}
  C.~Giunti and M.~Tanimoto,
  Phys.\ Rev.\ D {\bf 66} (2002) 113006. 

\bibitem{qlc}
  H.~Minakata and A.~Y.~Smirnov,
  Phys.\ Rev.\ D {\bf 70} (2004) 073009. 

\bibitem{raidal}
M.~Raidal,
Phys.\ Rev.\ Lett.\  {\bf 93} (2004) 161801. 


\bibitem{frampton} 
  D.~A.~Eby and P.~H.~Frampton,
 Phys.\ Rev.\ D {\bf 86} (2012) 117304. 

\bibitem{myrev}  A.~Y.~Smirnov,
  J.\ Phys.\ Conf.\ Ser.\  {\bf 447} (2013) 012004
 [arXiv:1305.4827].  


\bibitem{weinberg} 
S.~Weinberg,
  Phys.\ Rev.\ Lett.\  {\bf 43} (1979) 1566.

\bibitem{seesaw}
 P.~Minkowski,
Phys.\ Lett.\ B {\bf 67} (1977) 421.
T. Yanagida, in Proc. of Workshop on Unified Theory and
Baryon number in the Universe, eds. O. Sawada and A. Sugamoto, KEK, Tsukuba, (1979); 
M. Gell-Mann, P. Ramond and R. Slansky, in Supergravity, eds. 
P. van Niewenhuizen and D. Z. Freedman (North Holland,
Amsterdam 1980); 
S. L. Glashow, in Quarks and Leptons, Cargese lectures, eds M. Levy, 
(Plenum, 1980, New York) p . 707; R. N. Mohapatra and G. Senjanovic,
Phys. Rev. Lett. {\bf 44}, (1980) 912.

\bibitem{vissani} 
  F.~Vissani,
  Phys.\ Rev.\ D {\bf 57} (1998) 7027; 
  M.~Farina, D.~Pappadopulo and A.~Strumia,
  JHEP {\bf 1308} (2013) 022.

\bibitem{stabil} 
  J.~Elias-Miro, J.~R.~Espinosa, G.~F.~Giudice, G.~Isidori, A.~Riotto and A.~Strumia,
  Phys.\ Lett.\ B {\bf 709} (2012) 222. 

\bibitem{dseesaw} D.~Wyler and L.~Wolfenstein,
  Nucl.\ Phys.\ B {\bf 218} (1983) 205.
 R.~N.~Mohapatra,
  Phys.\ Rev.\ Lett.\  {\bf 56} (1986) 561.
  R.~N.~Mohapatra and J. F. W. Valle,  Phys.\ Rev.\ D {\bf 34} (1986) 1642.

\bibitem{dev}
  P.~S.~B.~Dev and A.~Pilaftsis,
  Phys.\ Rev.\ D {\bf 86} (2012) 113001. 


\bibitem{add}
 N.~Arkani-Hamed, S.~Dimopoulos, G.~R.~Dvali and J.~March-Russell,
  Phys.\ Rev.\ D {\bf 65} (2002) 024032. 

\bibitem{rs}
 Y.~Grossman and M.~Neubert,
  Phys.\ Lett.\ B {\bf 474} (2000) 361. 


\bibitem{1loop} 
  E.~Ma,
  Phys.\ Rev.\ D {\bf 73} (2006) 077301
  [hep-ph/0601225].

\bibitem{babu} K.~S.~Babu,
  Phys.\ Lett.\ B {\bf 203} (1988) 132, 
 A.~Zee,
  Phys.\ Lett.\ B {\bf 93} (1980) 389
   [Erratum-ibid.\ B {\bf 95} (1980) 461].
  K.~S.~Babu, A.~Patra and S.~K.~Rai,
  Phys.\ Rev.\ D {\bf 88}, 055006 (2013). 


\bibitem{LRmod}  A.~Maiezza, M.~Nemevsek, F.~Nesti and G.~Senjanovic,
  Phys.\ Rev.\ D {\bf 82} (2010) 055022;  
M.~Nemevsek, F.~Nesti, G.~Senjanovic and Y.~Zhang,
  Phys.\ Rev.\ D {\bf 83} (2011) 115014; 
  P.~S.~B.~Dev and R.~N.~Mohapatra,
  arXiv:1308.2151 [hep-ph].

\bibitem{goran} W.~-Y.~Keung and G.~Senjanovic,
  Phys.\ Rev.\ Lett.\  {\bf 50} (1983) 1427.

\bibitem{lhcbb}
 V.~Tello, et al, 
  Phys.\ Rev.\ Lett.\  {\bf 106} (2011) 151801, 
P.~S. Bhupal Dev, et al, 1305.0056 [hep-ph]. 

\bibitem{numsm}
  T.~Asaka, S.~Blanchet and M.~Shaposhnikov,
  Phys.\ Lett.\ B {\bf 631} (2005) 151, 
  L.~Canetti, M.~Drewes, T.~Frossard and M.~Shaposhnikov,
  Phys.\ Rev.\ D {\bf 87} (2013) 093006. 

\bibitem{bezrukov}
  F.~L.~Bezrukov and M.~Shaposhnikov,
  Phys.\ Lett.\ B {\bf 659} (2008) 703.


\bibitem{tbm}
  P.~F.~Harrison, D.~H.~Perkins and W.~G.~Scott,
  Phys.\ Lett.\ B {\bf 530} (2002) 167, 
 L.~Wolfenstein,
  Phys.\ Rev.\ D {\bf 18} (1978) 958.


\bibitem{discrete}
S.~Pakvasa and H.~Sugawara,
  Phys.\ Lett.\ B {\bf 73} (1978) 61.
 G.~C.~Branco,
  Phys.\ Lett.\ B {\bf 76} (1978) 70, 
  E.~Ma and G.~Rajasekaran,
  Phys.\ Rev.\ D {\bf 64} (2001) 113012. 

\bibitem{framework} 
C.~S.~Lam,
  Phys.\ Rev.\ D {\bf 78} (2008) 073015, 
  Phys.\ Rev.\ Lett.\  {\bf 101} (2008) 121602, 
 C.~S.~Lam,
  Phys.\ Rev.\ D {\bf 87} (2013) 013001, 
  W.~Grimus, L.~Lavoura and P.~O.~Ludl,
  J.\ Phys.\ G {\bf 36} (2009) 115007, 
  W.~Grimus and L.~Lavoura,
  JHEP {\bf 0904} (2009) 013. 

\bibitem{dani1} 
 D.~Hernandez and A.~Y.~Smirnov,
 Phys.\ Rev.\ D {\bf 86} (2012) 053014;  
 Phys.\ Rev.\ D {\bf 87} (2013) 5,  053005. 

\bibitem{Lim}
  M.~Holthausen, K.~S.~Lim and M.~Lindner,
  Phys.\ Lett.\ B {\bf 721} (2013) 61. 


\bibitem{lam-2}
C.~S.~Lam,
  Phys.\ Rev.\ D {\bf 87} (2013) 053018. 

\bibitem{dani3}
  D.~Hernandez and A.~Y.~Smirnov,
  Phys.\ Rev.\ D {\bf 88} (2013) 093007. 

\bibitem{joshipura}
 A.~S.~Joshipura and K.~M.~Patel,
  Phys.\ Lett.\ B {\bf 727} (2013) 480. 



\bibitem{giunti} C. Giunti, {\it these proceedings}.

\bibitem{renata} 
  A.~Y.~Smirnov and R.~Zukanovich Funchal,
  Phys.\ Rev.\ D {\bf 74} (2006) 013001. 




\end{thebibliography}
\end{document}